\documentclass[10pt]{article}
\usepackage{graphicx}
\usepackage{amssymb}
\usepackage{amsmath}
\usepackage{multicol,multirow}
\usepackage{epsf,psfrag,pstricks}
\usepackage{rotating}
\usepackage{algorithm,algorithmicx}
\usepackage[noend]{algpseudocode}
\renewcommand\algorithmicensure{\textbf{Output:}}
\renewcommand\algorithmicrequire{\textbf{Input:}}
\makeatletter
\newcommand\figcaption{\def\@captype{figure}\caption}
\newcommand\tabcaption{\def\@captype{table}\caption}
\makeatother
\begin{document}
\title{A Closed-Form Method for LRU Replacement under Generalized Power-Law Demand\thanks{Work conducted at Boston University with support from a Marie Curie
Outgoing International Fellowship of the EU
(MOIF-CT-2005-007230).} }
\author{Nikolaos Laoutaris\\
\texttt{nlaout@eecs.harvard.edu} \vspace{10pt}
\\ Division of Engineering and Applied Sciences, Harvard University \\33 Oxford Street, Cambridge, MA 02138}
\date{}
\maketitle
\begin{abstract}
We consider the well known \emph{Least Recently Used} (LRU)
replacement algorithm and analyze it under the independent
reference model and generalized power-law demand. For this
extensive family of demand distributions we derive a closed-form
expression for the per object steady-state hit ratio. To the best
of our knowledge, this is the first analytic derivation of the per
object hit ratio of LRU that can be obtained in constant time
without requiring laborious numeric computations or simulation.
Since most applications of replacement algorithms include (at
least) some scenarios under i.i.d. requests, our method has
substantial practical value, especially when having to analyze
multiple caches, where existing numeric methods and simulation
become too time consuming.
\end{abstract}
\section{Introduction}
Although very simple in both conception and implementation, the
LRU replacement algorithm is notoriously hard in terms of
analysis. Attempts to obtain the per object steady-state hit ratio
in an LRU operated cache under the independent reference model
(IRM)~\cite{coffman73} date back to the early 70's and have
continued appearing in the literature until very
recently~\cite{Che:Filtering,Laoutaris2005:PEVA,panagakis2005:LRU}.
As elaborated later on in this article, such attempts yield either
(1) intractable numeric methods for obtaining the exact hit
probabilities~\cite{King1971:Analysis,coffman73,Starobinski2000:Probabilistic},
(2) tractable numeric methods for obtaining approximate hit
probabilities~\cite{Flajolet1992:Birthday,Dan1990:Approximate,Che:Filtering,Laoutaris2005:PEVA,panagakis2005:LRU},
or (3) asymptotic results under infinite number of objects and
infinite storage
capacity~\cite{jalenkovic1998:Asymptotic,jalenkovic2003:Asymptotic}.
In this article we derive for the first time a closed-form formula
that can be used for obtaining approximate hit probabilities in
constant time, i.e., without numeric computation that depends on
input parameters like the number of objects and the storage
capacity. Although previous approximate numeric methods are fast
(linear complexity), being able to compute the hit probabilities
in constant time gives a significant advantage, especially when
the object universe is large or when there are more than one
caches to be analyzed. Examples include networks of
inter-connected cooperative
caches~\cite{Rodriguez2001:Analysis,Korupolu2002:Coordinated,Laoutaris2006:Mistreatment},
peer-to-peer caching systems~\cite{Xiao2004:P2Pcache}, semantic
caching and query processing~\cite{Ren2003:semanticcaching}.

We achieve the aforementioned result for generalized power-law
demand
distributions~\cite{Mitzenmacher2003:powerlaw,faloutsos1996:multifractals}.
Our interest on this family is based on the fact that such
popularity profiles have been observed in many real-world
measurement studies related to replacement algorithms,
including~\cite{Breslau1999:Web,Mahanti2000:Traffic}. It is also
quite a versatile family as it includes a wide range of profiles,
from uniform (having skewness parameter $a=0$) to Zipf (having
skewness parameter $a=1$). Most new applications that include a
cache\footnote{\footnotesize{We would like at this point to
emphasize the distinction between caching systems and replacement
algorithms. A caching system involves many more design choices
other than the particular replacement algorithm (there are issues
of associativity, multi-level hierarchical structure, and others).
The current article is about analyzing a particular replacement
algorithm and does not make any claims about the more general
problem of designing cache memories.}} that operates under LRU
replacement, typically include among others, experimental results
under power-law popularity; in these cases our closed-form method
can be used instead of laborious numeric methods or simulation.
\section{Related Work}
The problem of analyzing the hit ratio of LRU can be traced back
to the 70's. King~\cite{King1971:Analysis} was the first to derive
the steady-state behavior of LRU under IRM. Initial attempts
employed a Markov chain to model the contents of a cache operating
under LRU. Unfortunately, such attempts give rise to huge Markov
chains, having $C! {N \choose C}$ states (where $N$ denotes the
total number of distinct objects, and $C$ denotes the capacity of
the cache in unit-sized objects); numerical results for such
chains can only be derived for very small $N$ and $C$. More
efficient steady-state formulas have been derived by avoiding the
use of Markov chains, and instead making combinatorial arguments;
see Koffman and Denning~\cite{coffman73}, and Starobinski and
Tse~\cite{Starobinski2000:Probabilistic}. However, such approaches
still incur a computational complexity that is exponential in $N$
and $C$. Flajolet et al.~\cite{Flajolet1992:Birthday} have
presented integral expressions for the hit ratio, which can be
approximated using numerical integration at complexity $O(NC)$.
Dan and Towsley~\cite{Dan1990:Approximate} have derived an $O(NC)$
iterative method for the approximation of the hit ratio.
Jalenkovi\'{c}~\cite{jalenkovic1998:Asymptotic} has provided a
closed form expression for the particular case of generalized
power-law demand with skewness parameter $\alpha>1$, for the
asymptotic case, $N,C\rightarrow \infty$. The same author has
shown that the hit ratio of LRU under such demand is
asymptotically insensitive for large caches, i.e., $C\rightarrow
\infty$, to temporal correlations of the request arrival
process~\cite{jalenkovic2003:Asymptotic}. The most recent attempts
on the analysis of LRU can be found
in~\cite{Che:Filtering,Laoutaris2005:PEVA,panagakis2005:LRU}.
These works build on the notion of \emph{characteristic time},
which is also used in our work. More details on these works and
the concept of the characteristic time are given in the following
sections.
\section{Background and Scope}
\begin{figure}[tb]
\psfrag{characteristic time r_i (units of
time)}{\Huge{characteristic time $r_i$ (units of time)}}
\centering
\includegraphics[angle=-90,width=4in]{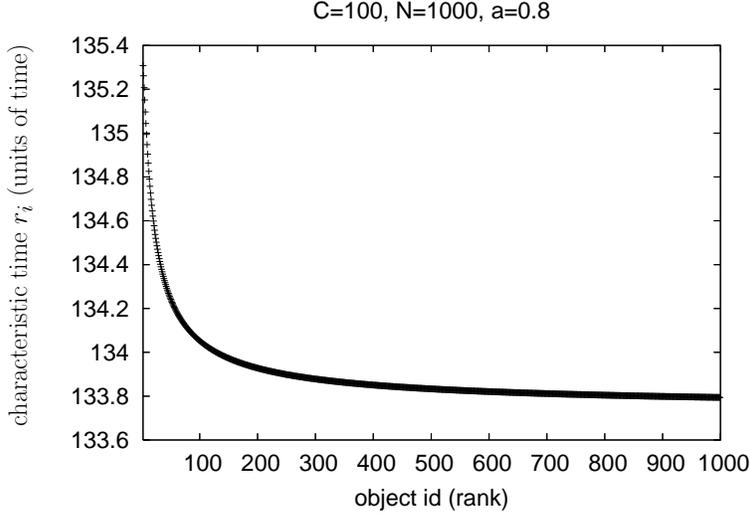} \caption{\footnotesize{An example with the characteristic times calculated by Eq.~(\ref{eq:haotau})}} \label{fig:notmuchdifferent}
\end{figure}

Consider an object set $O=\{o_1,\ldots,o_N\}$, where $o_i$ denotes
the $i$th unit-sized object. Assume that requests are issued for
the objects of $O$ and that successive requests are
independent\footnote{\footnotesize{The independent reference
model~\cite{coffman73} is commonly used to characterize cache
access patterns~\cite{arlitt:sigmetrics96,Breslau1999:Web}. The
impact of temporal correlations was shown in
\cite{JinBestavros:mascotts,Psounis2004:Correlations} to be
minuscule, especially under typical, Zipf-like object popularity
profiles. These works showed that temporal correlations decrease
rapidly with the distance between any two samples so, as long as
the cache size is not minuscule, they do not impact fundamentally
on the i.i.d. assumption. The unit assumption regarding the size
of objects is a standard one in all previous
works~\cite{coffman73}--\cite{jalenkovic2003:Asymptotic} and stems
from the desire to avoid adding 0/1-knapsack type complexities to
a problem that is already combinatorial. Practically, it is
justified on the basis that in many caching systems the objects
are much smaller than the available cache size. Similarly, all
previous works assume stationarity of demand over some time
horizon. This is supported by many of the aforementioned
measurement works, over multiple time scales. Obviously, if the
demand is non stationary and radically changing over small time
scales, no analysis can be carried out.}} and identically
distributed according to a common probability distribution
$\vec{p}=\{p_1,\ldots,p_N\}$, where $p_i$ denotes the request
probability for the $i$th most popular object of $O$ (hereafter
assumed to be object $o_i$ without loss of generality). The
aggregate stream of requests is assumed to be arriving to a cache
according to a Poisson arrival process\footnote{\footnotesize{We
can alternatively obtain similar results by assuming a Bernoulli
arrival process and carrying-out a discrete time analysis. We
choose to remain on the continuous time domain so as to be aligned
with the preceding body of work
in~\cite{Che:Filtering,Laoutaris2005:PEVA,panagakis2005:LRU}.}} of
rate $\lambda$~requests/unit of time (meaning that the stream of
request for any given object is also Poisson with rate $\lambda
\cdot p_i, 1\leq i\leq N$). In Laoutaris et
al.~\cite{Laoutaris2005:PEVA} we showed that under the above
mentioned request model, an LRU operated cache with capacity for
$C$ unit-sized objects reaches a steady-state in which the
probability of finding object $o_i$ in the cache is given by:
\begin{equation}\label{eq:hitratio}
\pi_i=1-e^{-p_i r_i}
\end{equation}
In the above equation, $r_i$ denotes the maximum inter-arrival
time between two adjacent request for object $o_i$, both of which
lead to hits. This quantity is referred to as the
\emph{characteristic time} of object $o_i$ and is due to Che et
al.~\cite{Che:Filtering}. In essence, $r_i$ is a random variable,
but it can be approximated by a constant in order to carry-out a
tractable analysis. This is characterized as a \emph{mean field
approximation} in~\cite{Che:Filtering} and the rationale behind it
is that when the object set is large enough, $r_i$ fluctuates
closely around its mean value, so it can be effectively approximated
by it. The characteristic time $r_i$ of object $o_i, 1\leq i\leq
N$, was obtained in~\cite{Che:Filtering,Laoutaris2005:PEVA} by
solving the following equation numerically:
\begin{equation}\label{eq:haotau}
\sum_{{j=1 \atop j\neq i}}^N 1-e^{-p_j r_i}=C \Rightarrow
\sum_{{j=1 \atop j\neq i}}^N e^{-p_j r_i}=N-1-C
\end{equation}
This equation gives the time interval that is required for the
$N-1$ other objects to generate $C$ distinct
requests\footnote{\footnotesize{Observe that the quantity within
the summation is the CDF of the exponential request inter-arrival
time for object $o_j$ calculated at point $r_i$.}} and thus evict
$o_i$, granted that $o_i$ is not re-requested in this interval.
However, solving $N$ such equations, one for each object, is
cumbersome, especially for large $N$. This can be partially
alleviated by considering a single characteristic time $r$ for all
the objects and thus solving only one equation. Such an
approximation is justifiable on the basis that the characteristic
times $r_i$ of different objects do not differ substantially, even
under skewed popularity distributions.
Figure~\ref{fig:notmuchdifferent} supports this claim by
illustrating the characteristic times of objects in an LRU cache
with capacity for $C=100$ objects that is driven by requests over
an object universe of $N=1000$ objects, whose popularities follow
a generalized power-law with skewness $a=0.8$ (the request rate
for this and all subsequent examples is normalized to
$\lambda=1$~request/unit of time). The characteristic times are
obtained by solving Eq.~(\ref{eq:haotau}) numerically. One can
observe that although $\vec{p}$  is skewed, the difference between
the characteristic times of different objects is very small (thus
$r_{1}/r_{1000}=1.011$ despite that $p_{1}/p_{1000}=251$, i.e.,
two orders of magnitude apart). The plot essentially says that
request inter-arrivals for the same object that are longer than
134-135 time units lead to misses.

The approach of using a common characteristic time $r$ for all
objects was recently employed by Panagakis et al.
in~\cite{panagakis2005:LRU}. In the same work it was observed that
the most natural way of finding the common characteristic time is by
solving the following \emph{normalization equation} which simply
requires that all the steady-state object hit probabilities sum up
to the capacity of the cache, i.e.:
\begin{equation}\label{eq:apantau}
\sum_{i=1}^N \pi_i=C \Rightarrow \sum_{i=1}^N 1-e^{-p_i
r}=C\Rightarrow \sum_{i=1}^N e^{-p_i r}=N-C
\end{equation}
The above equation was solved numerically
in~\cite{panagakis2005:LRU}, similarly to the case
of~\cite{Che:Filtering,Laoutaris2005:PEVA} and
Eq.~(\ref{eq:haotau}). In the following section, we utilize the
notion of characteristic time as developed
in~\cite{Che:Filtering,Laoutaris2005:PEVA,panagakis2005:LRU} and
present an analysis that leads to the derivation of a closed-form
formula for the behavior of LRU caching. This is, to the best of
our knowledge, the first, non-asymptotic, closed-form approximate
formula for LRU (the closed-form expression of Jalenkovi\'{c} in
\cite{jalenkovic1998:Asymptotic} covers only the asymptotic case
$N,C \rightarrow \infty$ and is only for $a>1$). Our method can be
used for the study of LRU caching, whether in stand-alone mode (a
single LRU cache), or, more interestingly, in
hierarchical~\cite{Che:Filtering,Laoutaris2005:PEVA} or
distributed~\cite{Laoutaris2006:Mistreatment} inter-connections of
caches, without requiring laborious numeric computations.
\section{Analysis of LRU under Generalized Power-Law Demand}
We assume that $\vec{p}$ follows a generalized
power-law distribution, in which the $i$th most popular object has
request probability $p_i=\Lambda/i^a$, where
$\Lambda=(\sum_{i'=1}^N \frac{1}{{i'}^a})^{-1}$ is a normalization
constant, and $a$ is a skewness parameter. Under such demand, we
show how to obtain an approximate closed-form formula for the
common characteristic time $r$ of Eq.~(\ref{eq:apantau}). This
gives directly a closed-form expression for the hit ratio of each
object through Eq.~(\ref{eq:hitratio}). Our analysis can be easily
adapted to handling per object characteristic times $r_i$. The
only difference in this case would be that we would start from
Eq.~(\ref{eq:haotau}) instead of Eq.~(\ref{eq:apantau}).

First we take the Taylor series expansion of the exponential form
$e^{-p_i r}$ in terms of the variable $r$ around point $C$:
\begin{equation}\label{eq:firsttaylor}
e^{-p_i r}=e^{-p_i C} \cdot \sum_{k=0}^\infty
\frac{\left(-p_i\cdot (r-C)\right)^k}{k!}
\end{equation}
The exponential form $e^{-p_i C}$ of Eq.~(\ref{eq:firsttaylor})
can be similarly expanded in terms of the variable $p_i$ around point $0$
as follows:
\begin{equation}\label{eq:secondtaylor}
e^{-p_i C}=\sum_{k=0}^\infty \frac{(-p_i C)^k}{k!}
\end{equation}
Using Eqs~(\ref{eq:firsttaylor}),~(\ref{eq:secondtaylor}) in
Eq.~(\ref{eq:apantau}) we can write:
\begin{equation}\label{eq:trans1}
\begin{array}{ll}
\displaystyle \sum_{i=1}^N e^{-p_i r}=N-C \Rightarrow \sum_{i=1}^N
\left( \sum_{k=0}^\infty \frac{(-p_i C)^k}{k!}\right)\cdot \left(
\sum_{k=0}^\infty \frac{\left(-p_i\cdot (r-C)\right)^k}{k!}
\right) =N-C
\end{array}
\end{equation}
Denoting $a_k=(-C)^k/k!$ and $b_k=\left( -(r-C)\right)^k/k!$, and
limiting $k$ to $0\leq k< K$ instead of letting it run to
$\infty$, we can approximate Eq.~(\ref{eq:trans1}) as follows:
\begin{equation}\label{eq:trans2}
\begin{array}{ll}
& \displaystyle \sum_{i=1}^N \left( \sum_{k=0}^K p_i^k \cdot a_k
\right) \cdot \left( \sum_{k=0}^K p_i^k \cdot b_k \right)
=N-C \Rightarrow \\
& \displaystyle \sum_{i=1}^N \left( \sum_{m=0}^{2K} p_i^m \cdot
\left( \sum_{{m_1,m_2: \atop {m_1\leq K, m_2\leq K \atop
m_1+m_2=m}}}a_{m_1}\cdot b_{m_2} \right) \right)
=N-C \Rightarrow \\
& \displaystyle \sum_{m=0}^{2K} \left( \left( \sum_{{m_1,m_2:
\atop {m_1\leq K, m_2\leq K \atop m_1+m_2=m}}}a_{m_1}\cdot b_{m_2}
\right) \cdot \left( \sum_{i=1}^N p_i^m \right)  \right) =N-C
\end{array}
\end{equation}
As will be shown later through numeric examples, the truncation to
$K$ has a small effect on the accuracy as compared to solving
Eq.~(\ref{eq:trans2}) for $K\rightarrow \infty$. This owes to the
fact that the remainder for $k>K$ of the previous exponential
forms~(\ref{eq:firsttaylor}),~(\ref{eq:secondtaylor}) can be
bounded by $O(1/K!)$.

We continue the analysis by putting into use our assumption that
$p_i$ follows a power-law distribution, and so we can write:
\begin{equation}\label{eq:trans3}
\sum_{i=1}^N p_i^m=\sum_{i=1}^N \left( \frac{\Lambda}{i^a}
\right)^m=\Lambda^m \cdot \sum_{i=1}^N \frac{1}{i^{am}}=\Lambda^m
\cdot  H_N^{(am)},
\end{equation}
where $H_N^{(a)}=\sum_{l=1}^N 1/l^a$ denotes the $N$th Harmonic
number of order $a$. $H_N^{(a)}$ can be approximated by its
integral expression $H_N^{(a)}\approx \frac{N^{1-a}-1}{1-a}$ (see
also~\cite{Tang2004:Hash}). Substituting from
Eq.~(\ref{eq:trans3}) into Eq.~(\ref{eq:trans2}) we obtain our
\emph{master equation}:
\begin{equation}\label{eq:master}
\sum_{m=0}^{2K} \left( \left( \sum_{{m_1,m_2: \atop {m_1\leq K,
m_2\leq K \atop m_1+m_2=m}}}a_{m_1}\cdot b_{m_2} \right) \cdot
\Lambda^m \cdot  H_N^{(am)}  \right) =N-C
\end{equation}
The master equation is a $K$-order polynomial equation of $r$
(corresponding to an approximate version of Eq.~(\ref{eq:trans1})
that retains only $K+1$ first terms from the Taylor series
expansions of the exponential forms of
Eqs~(\ref{eq:firsttaylor}),~(\ref{eq:secondtaylor})). One can
solve the master equation in arbitrary accuracy by increasing $K$.
This, of course, presumes a numerical solution and, thus, does not
differ fundamentally from the previous numerical approaches
in~\cite{Che:Filtering,Laoutaris2005:PEVA,panagakis2005:LRU}.
Where the master equation is essentially different, is in that it
has a form that can be utilized for setting up a closed-form
solution. This can be accomplished by selecting appropriately
small $K$ that give rise to such results. Such flexibility is not
provided by Eqs.~(\ref{eq:haotau}),~(\ref{eq:apantau}).

Consider the case of $K=2$. Substituting $a_m, b_m$ and doing some
algebraic manipulation reduces the master equation into the
following quadratic equation ($K=2$ amounts to retaining the first
three terms of the Taylor series expansions of
Eqs~(\ref{eq:firsttaylor}),~(\ref{eq:secondtaylor})):
\begin{equation}\label{eq:quadratic}
\begin{array}{ll}
\displaystyle \alpha_2 r^2 + \alpha_1 r+\alpha_0=0 \qquad
\mbox{where:} & \displaystyle \alpha_2= \frac{\Lambda^2}{2}
H_N^{(2a)}-\frac{\Lambda^3C}{2} H_N^{(3a)}+\frac{\Lambda^4C^2}{4}
H_N^{(4a)} \vspace{10pt} \\ & \displaystyle \alpha_1= -\Lambda
H_N^{(a)}+\frac{\Lambda^3C^2}{2} H_N^{(3a)}-\frac{\Lambda^4C^3}{2}
H_N^{(4a)} \vspace{10pt} \\ & \displaystyle
\alpha_0=C+\frac{\Lambda^4C^4}{4} H_N^{(4a)}
\end{array}
\end{equation}
The characteristic time can then be taken by selecting an
appropriate real solution (assuming that one exists, more on this
in the sequel) from the quadratic formula: $r=\frac{-\alpha_1\pm
\sqrt{\alpha_1^2-4\alpha_2 \alpha_0 }}{2\alpha_2}$.

We can go a step further and consider the case of $K=3$ which
yields the following cubic equation:
\begin{equation}\label{eq:cubic}
\begin{array}{ll}
\displaystyle \alpha_3 r^3 +\alpha_2 r^2 + \alpha_1 r+\alpha_0=0
\qquad \mbox{where:} & \displaystyle \alpha_3=-\frac{\Lambda^3}{6}
H_N^{(3a)} + \frac{\Lambda^4 C}{6} H_N^{(4a)} - \frac{\Lambda^5
C^2 }{12} H_N^{(5a)}+\frac{\Lambda^6 C^3}{36} H_N^{(6a)}
\vspace{10pt}
\\& \displaystyle \alpha_2= \frac{\Lambda^2}{2}
H_N^{(2a)}-\frac{\Lambda^4 C^2}{4} H_N^{(4a)} + \frac{\Lambda^5
C^3}{6} H_N^{(5a)}-\frac{\Lambda^6 C^4}{12}
H_N^{(6a)}\vspace{10pt}
 \\ & \displaystyle \alpha_1= -\Lambda H_N^{(a)}+\frac{\Lambda^4 C^3}{6} H_N^{(4a)}-\frac{\Lambda^5 C^4}{12} H_N^{(5a)}+\frac{\Lambda^6 C^5}{12} H_N^{(6a)} \vspace{10pt} \\ & \displaystyle
\alpha_0=C-\frac{\Lambda^4 C^4}{12} H_N^{(4a)}-\frac{\Lambda^6
C^6}{36} H_N^{(6a)}
\end{array}
\end{equation}
The cubic formula~\cite{Nickalls:cubic} (we do not repeat it here
due to space considerations) returns the three solutions to the
above cubic equation expressed as analytic functions of the
coefficients $\alpha_3,\alpha_2,\alpha_1,\alpha_0$ (which, in
turn, are analytic functions\footnote{\footnotesize{For the
generalized Harmonic number we use its integral approximation as
stated earlier on.}} of the input parameters $C,N,a$); at least
one the three solutions is always guaranteed to be in the domain
of real numbers (such a guarantee does not exist for the quadratic
equation, for which, both solutions can be complex). Due to this
guarantee, and also to the fact that it provides a closer
approximation by considering an additional term from the Taylor
expansion, we focus on the $K=3$
case.\footnote{\footnotesize{Theoretically we could go even
further and consider the quartic equation ($K=4$). This, however,
involves very cumbersome formulas for the roots and is marginally
valuable since the cubic equation already provides close
approximation as will be demonstrated in Sect.~\ref{sec:numerics}.
The quintic and all higher order equations ($K\geq 5$) do not
posses a general solution over the rationals in terms of radicals
(the ``Abel-Ruffini'' theorem).}} Let $r_A, r_B, r_{\Gamma}$ be
the three roots of Eq.~(\ref{eq:cubic}) returned by the cubic
formula. We select as characteristic time the smallest real
solution $r_X, X\in \{A,B,\Gamma\}$ that exceeds $C$, i.e.:
\begin{equation}\label{eq:rcubic}
r=\min_{X\in\{A,B,\Gamma\}} \left( r_X \right): r_X\in\mathbb{R},
r_X \geq C
\end{equation}
The rationale behind this choice is that it takes at least $C$
requests to evict a newly inserted object so the characteristic
time has to be larger than $C$ (the characteristic time is in
units of time or alternatively in number of requests, since we
have normalized the request rate $\lambda$ into 1 req./time slot).
In the next section we show that the above approximation yields
accurate $r$ and $\pi_i$ across a wide range of parameters
$C,N,a$.
\section{Numeric Results}\label{sec:numerics}
In this section we first compare the accuracy of the approximate
characteristic time that we obtain from Eq.~(\ref{eq:rcubic}) with
the exact characteristic time that we obtain from solving
Eq.~(\ref{eq:apantau}) numerically. Table~\ref{table:chartimes}
provides such a comparison drawn from a universe of $N=1000$
objects and for varying $a$ and $C$. Each cell of the table
corresponds to an $(a,C)$ pair and contains two numeric values:
the top one is the exact characteristic time while the bottom one
is the approximate one that we compute through our method. These
values correspond to units of time, or equivalently, number of
requests.

One may observe that our approximation tracks closely the actual
characteristic time. Deviations appear only under very skewed
demand (e.g., $a\geq0.8$) and large relative storage capacities
(e.g., $C/N\geq 20\%$). These cases, however, are neither typical,
nor really interesting, for the following reasons. First, cache
memories rarely operate under so much storage. Typical values for
the ratio $C/N$ are well below $10\%$ in most applications (this
is after all the main reason for employing caches -- lack of
memory space for all the objects). Second, a high availability of
storage, combined with a high skewness, leads to a fairly expected
cache hit ratio that approaches 1 and, thus, there is not much
practical purpose for studying such a case analytically. We note,
however, that our method can be twicked in order to provide useful
results for these cases also. We show how to do this later in this
section.
\begin{table}[tb]
\centering
\begin{tabular}{c|cccc}
  $a\backslash C$ & 50 & 100 & 150 & 200 \\ \hline
  \multirow{2}{*} {0.4} & 51.8 & 107.5 & 167.5 & 232.2 \\ & 52 &
  107.9 & 167.8 & 232.1 \\ \hline
  \multirow{2}{*} {0.6} & 53.6 & 114.3 & 181.9 & 256.7 \\ & 54 &
  113.9 & 178.6 & 248.9 \\ \hline
  \multirow{2}{*} {0.8} & 59.6 & 133.8 & 220.2 & 318.6 \\ & 59.1 &
  128.9 & 167.5 & 225.2
\end{tabular}
\caption{\footnotesize{Evaluation of the accuracy of our
approximate closed-form formula for the characteristic time on a
set of $N=1000$ objects, under varying cache size $C$ and demand
skewness $a$. The top value of each cell gives the exact
characteristic time from solving Eq.~(\ref{eq:apantau})
numerically while the bottom value gives the approximate
characteristic time from
Eq.~(\ref{eq:rcubic}).}}\label{table:chartimes}
\end{table}

\begin{figure}[tb]
\centering
\includegraphics[angle=-90,width=4in]{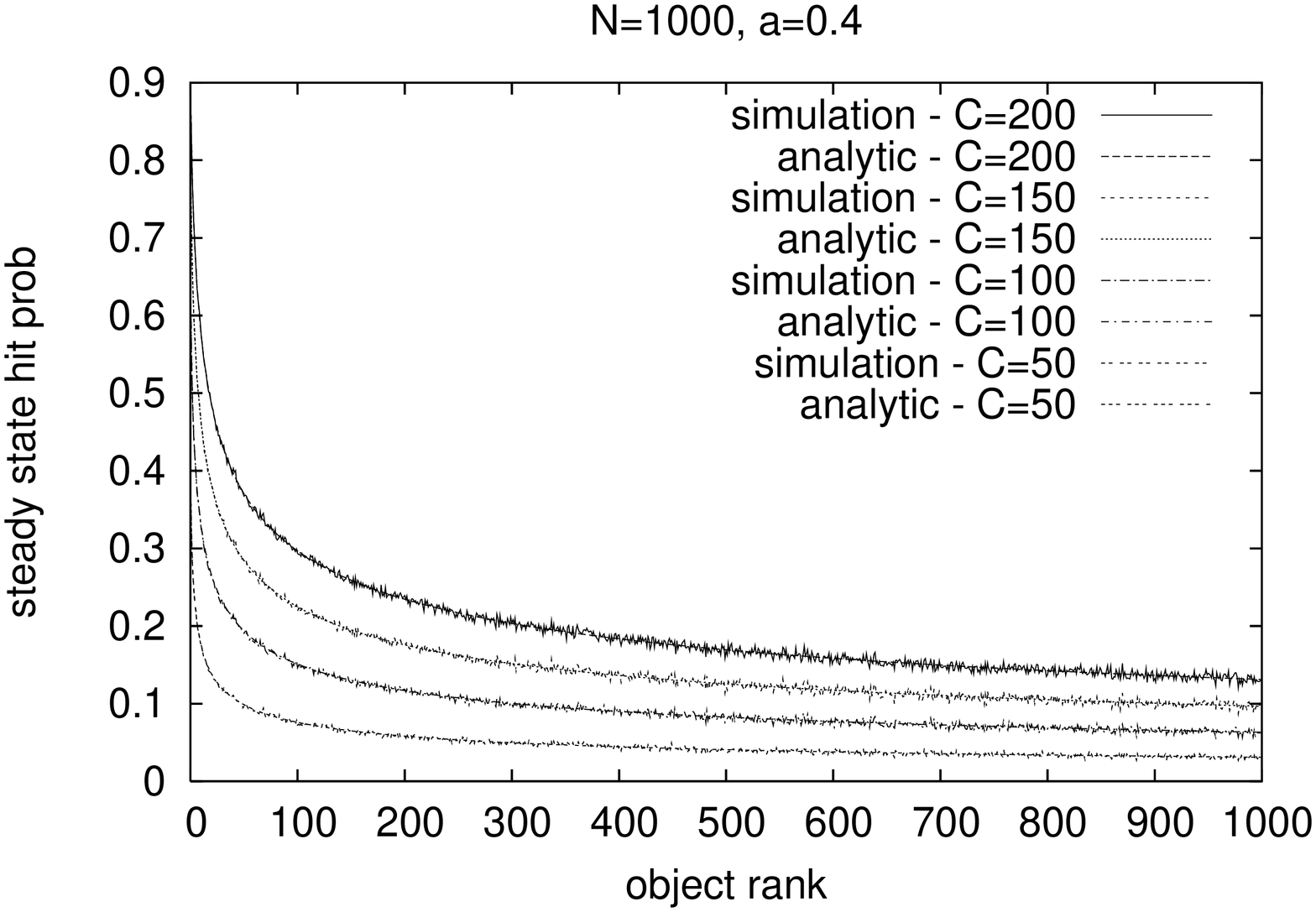}
\includegraphics[angle=-90,width=4in]{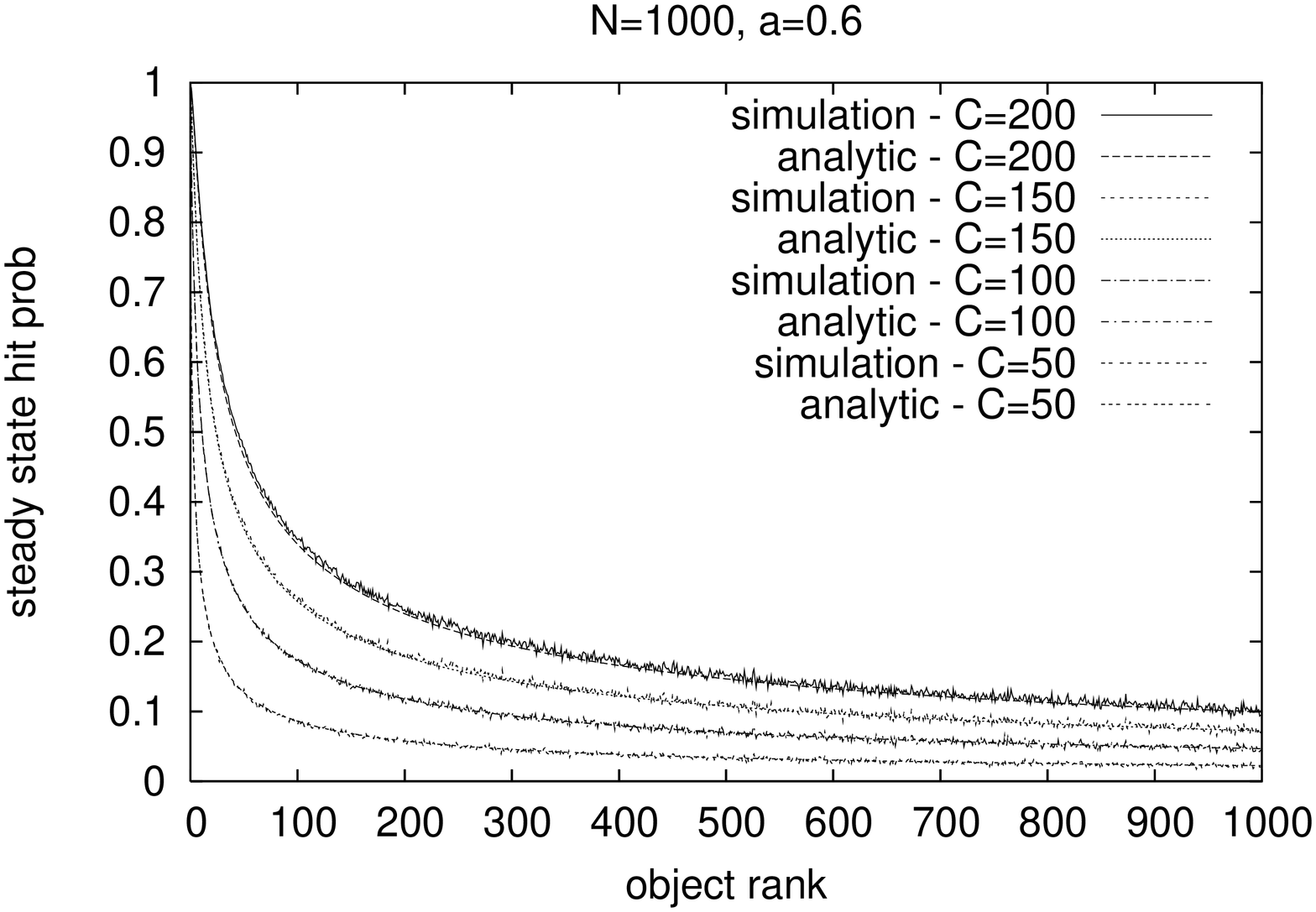}
\includegraphics[angle=-90,width=4in]{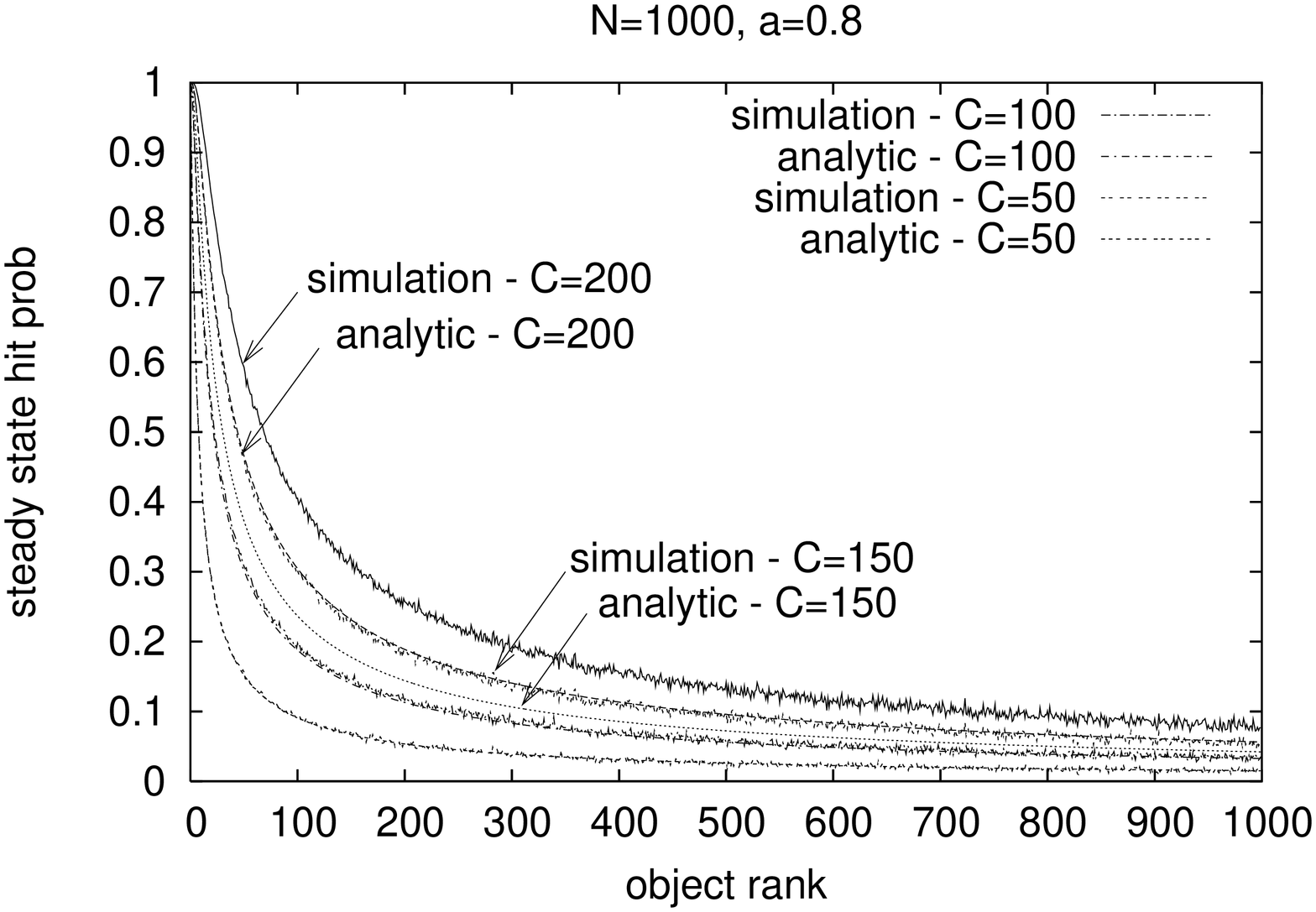}
\caption{\footnotesize{Comparison of simulated and analytic per
object hit probabilities ($\pi_i$'s) on a universe of $N=1000$
objects for different storage capacities ($C=50,100,150,200$) and
skewness parameters ($a=0.4,0.6,0.8$) for the input generalized
power-law demand. A word of caution: in the third graph ($a=0.8$)
the analytic line for $C=200$ overlaps coincidentally with the
simulation line for $C=150$.}} \label{fig:hitratios}
\end{figure}

The next set of results compares the analytic per object
steady-state hit probabilities obtained by plugging the
characteristic time $r$ of Eq.~(\ref{eq:rcubic}) into
Eq.~(\ref{eq:hitratio}), with corresponding hit probabilities
obtained by simulating LRU for 10~million requests. The three
graphs of Fig.~\ref{fig:hitratios} correspond to skewness $a=0.4$,
0.6 and 0.8. Each graph includes 8 curves corresponding to results
obtained from simulation and analysis under different ratios
$C/N=5\%$, $10\%$, $15\%$ and $20\%$. One may observe that for low
($a=0.4$) and medium ($a=0.6$) skewness, the analytically computed
hit ratios match almost perfectly with the simulated ones, across
all storage availabilities. For high skewness ($a=0.8$), our
results are very accurate up to a storage availability of $10\%$
and then start to deviate (some deviation for $C/N=15\%$ and a
larger one for $C/N=20\%$). In other words, the method becomes
less accurate under very skewed demand and large availability of
storage. The reason for this deviation is that under such
settings, the omission of higher order terms of the Taylor series
expansion of the previously mentioned exponential forms, disrupts
significantly the balance of the (normalization)
Eq.~(\ref{eq:apantau}), thus leading to $\pi_i$'s that do not sum
up to $C$. As we commented earlier, a storage availability higher
than $10\%$ is not realistic under most caching applications.
Nevertheless, in the following paragraph we will describe
\emph{proportional normalization}, a method for fixing this
problem by reshaping the $\pi_i$'s and, actually, achieving a high
accuracy even under high storage availability and skewed demand.
\\

\begin{figure}[t]
\centering
\includegraphics[angle=-90,width=4in]{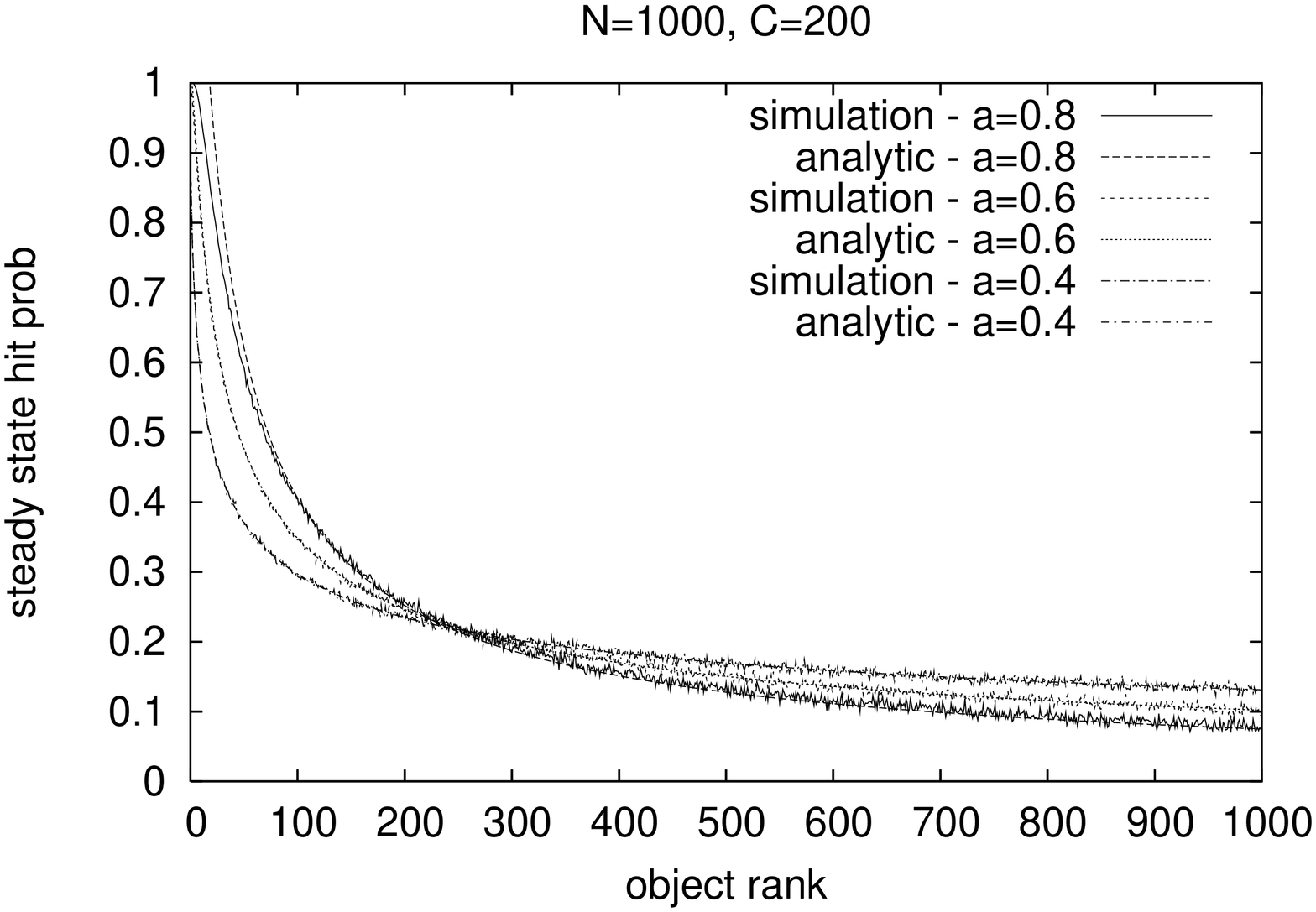}
\caption{\footnotesize{Comparison of simulated and proportionally
normalized analytic per object hit probabilities ($\pi_i$'s) on a
universe of $N=1000$ objects for a storage capacity $C=200$ and
different skewness parameters ($a=0.4,0.6,0.8$) for the input
generalized power-law demand.}} \label{fig:hitratiosfixed}
\end{figure}

\begin{algorithm}[tb]
\footnotesize \caption{ProportionalNormalization($\pi$: $1\times
N$ vector, $N,C$ scalars)}
\begin{algorithmic}[1]
\For{$i=1$ to $N$} \State $m\_mass=C-\sum_{j=1}^N \pi_j$; \State
$\delta = m\_mass \cdot \pi_i / \sum_{j=i}^N \pi_j$; \State
$\pi_i=\min\{\pi_i+\delta,1\}$; \EndFor
\end{algorithmic}
\label{alg:ProportionalNormalization}
\end{algorithm}
\noindent \textbf{Proportional normalization:} In this section we
describe a simple normalization method for fixing the missing
probability mass problem that occurs under combined high $C/N$ and
$a$. This is achieved through a proportional normalization method
that distributes the missing probability mass among the different
objects in such a way that each object's hit probability is
incremented proportionally to its hit probability as derived by
our base-line closed-form method. In
Algorithm~\ref{alg:ProportionalNormalization} we describe the
proportional normalization method. The algorithm takes as input
the vector of hit probabilities derived from
Eq.~(\ref{eq:hitratio}) after plugging in the analytically
computer characteristic time $r$ and returns a normalized vector
of hit probabilities that sum up to $C$. In
Fig.~\ref{fig:hitratiosfixed} we compare the normalized hit
probabilities with the corresponding ones from simulation under a
storage availability $C/N=20\%$ (under such availability, and for
high skewness, there was a substantial disagreement between
simulation and analysis, as shown in the third graph of
Fig.~\ref{fig:hitratios}). From Fig.~\ref{fig:hitratiosfixed} it
is clear that after the normalization there is almost perfect
agreement between the simulation and the analytic results. Thus by
combining our analytic method with proportional normalization, one
can obtain accurate hit ratios even under combined high storage
availability and skewed demand.
\section{Conclusions}
In this work we have presented a closed-form approximate method
for obtaining the per object hit ratio under LRU replacement and
independent generalized power-law requests. Our method obtains
accurate results for a wide range of parameters. It becomes less
accurate only when combining a very high storage availability
(which is not typical under most caching applications) with skewed
demand. To accommodate this case, we describe a simple
proportional normalization procedure that, when combined with our
baseline closed-form method, corrects its accuracy. To the best of
our knowledge, our method is the first one to produce non
asymptotic closed-form results for LRU. Due to the complete lack
of any kind of numeric computation our method can be used for the
analysis of large networks of LRU caches in which existing numeric
methods and simulation become impractical from a computational
point of view.

\bibliographystyle{../../../styles/IEEE}
\bibliography{../../../Nikos}

\end{document}